\documentclass[prb,preprint]{revtex4}
\usepackage{amsmath}
\usepackage{amsfonts}
\usepackage{amssymb}
\usepackage{graphicx}
\begin{document}

\title{Theoretical investigation of methane under pressure}
\author{Leonardo Spanu$^{1}$, Davide Donadio$^{1}$, Detlef Hohl$^{2}$, Giulia Galli$^{1}$}
\affiliation{$^{1}$ Department of Chemistry, University of California at Davis, Davis, CA\\
$^{2}$ Shell International Exploration and Production,   Rijswijk, The Netherlands}

\begin{abstract}

We present computer simulations of liquid and solid phases of condensed 
methane at pressures below 25 GPa, between 150 and 300 K,  where no 
appreciable molecular dissociation occurs.
We used  molecular dynamics (MD) and metadynamics techniques, and 
empirical potentials in the rigid molecule approximation, whose validity 
was confirmed a posteriori by carrying out {\it ab initio} MD simulations for 
selected pressure and temperature conditions. Our results for the 
melting line are in satisfactory agreement with existing measurements. 
We find that the fcc crystal transforms into a hcp structure with 4 
molecules per unit cell (B phase) at about 10 GPa and 150 K, and that 
the B phase transforms into a monoclinic high pressure phase above 20 
GPa. Our results for solid/solid phase transitions are consistent with 
those of Raman studies but the phase boundaries estimated in our 
calculations are at higher pressure than those inferred from 
spectroscopic data.

 \end{abstract}
\maketitle

\section{Introduction}
Methane is the most abundant organic molecule in the universe, and the phase diagram of its condensed forms is of great relevance in several branches of science, including planetary physics and petrology.  For example, spectroscopic data have revealed ~\cite{quirico} the presence of solid methane on the surface of different planets, e.g. Pluto ~\cite{plutoI,plutoII}, and the properties of methane at moderate pressures are crucial for modeling and understanding the formation and stability of hydrocarbons in the Earth's mantle and crust.
The phase diagram of methane is rather complex and poorly understood: at low pressure it exhibits several solid phases differing both for the positions of the carbon atoms and for the orientation of the molecules in the unit cell~\cite{Bini}.

At room temperature and $P=1.6$ GPa  methane crystallizes in the so called phase I, with C atoms occupying fcc lattice sites while H atoms are free to rotate ~\cite{hazen}; phase I has one molecule per unit cell. By isothermal compression at room temperature starting at $1.6$ GPa ~\cite{hazen}, transformations to different phases have been observed by X-Ray diffraction~\cite{hirai,nakahata,umemoto} 
IR and Raman studies~\cite{Bini}; these transitions occur approximately at $5$GPa (phase I to phase A), $12-18$ GPa ( A to B ) 
and $25$ GPa (B to a so called high-pressure phase-HP). Bini {\sl et al.}~\cite{Bini}, based on IR and Raman data, proposed a 
tetragonal crystal structure for the phase A, while X-ray diffraction data\cite{nakahata} 
indicate a rhombohedral structure.

Given the similarities between the electronic structure of the methane molecule (a closed shell  with 8 valence electrons) and rare gas atoms, Bini {\sl et al.} proposed 
that phase B has a hexagonal closed-packed (hcp) structure with one molecule per cell, similar to  that observed in diamond anvil-cell studies~\cite{hcp_goncharov} for   Xe~\cite{Xe}, Kr~\cite{Kr} and Ar ~\cite{Ar}. Umemoto~\cite{umemoto} and Hirai~\cite{hirai} proposed instead a cubic 
structure for phase B. All experiments found that the A-B transition is very sluggish, 
suggesting a possible, complex  structural rearrangement taking place between the two phases. Hirai {\sl et al.} also  
suggested the existence of a phase intermediate between A and B,
which they called pre-B, with a diffraction pattern totally different from the one of phase A. 
At pressure above $25$ GPa, a so called high pressure (HP) phase appears, and the B-HP 
transition is not accompanied by a change in the spatial arrangement of the carbon atoms.  Finally the melting line of methane has been measured between $100$ K and $450$ K in three different experiments ~\cite{kennedy,yagi, steyland}.

In this paper we present molecular dynamics (MD) simulations based on empirical potentials, aimed at investigating both the melting line of methane and its solid phases below 25 GPa and between $150$K and $300$K, i.e. in a regime where CH$_4$ molecules do not dissociate.  We used a metadynamics algorithm~\cite{metadynamics,metacell,natmat2006} to investigate solid/solid transitions and compared the results of several techniques in the case of the solid/liquid transition. In addition we performed {\it ab initio} MD simulations using Density Functional Theory (DFT) to validate a posteriori the empirical model utilized in our classical MD simulations. We do not consider $T<150$K, where quantum effects are expected to be important for the description of nuclear motion. 

The paper is organized as follows. In Section II we briefly describe some technical aspects of both classical and {\it ab initio} simulations. In Section III we present calculations of the melting line of methane, comparing results obtained by thermodynamic integration~\cite{frenkelti} with those of a two-phase coexistence method~\cite{twophase94,bonev}. In section IV we discuss metadynamics results for solid/solid phase transitions and  we justify a posteriori the use of the selected forcefield, by carrying out {\it ab initio} MD simulations for several (P,T) points. Finally in setion V we draw our conclusions.

\section{Computational Techniques}
We performed molecular dynamics (MD) simulations with the DLPOLY code \cite{dlpoly}, using the 
 empirical force field TraPPE-EH from Ref.~\cite{potenziale}, which was parametrized to describe vapor-liquid equilibria in alkanes. CH$_4$ molecules are treated as rigid bodies
and the intermolecular interactions are described by Lennard Jones (LJ) sites located on the
carbon atoms, and on the centers of carbon-hydrogen bonds~\cite{potenziale}. As we discuss in detail below, {\it ab initio} calculations confirmed a posteriori the validity of the rigid molecule approximation.
The MD equations of motions
were integrated using the leap-frog Verlet algorithm with a time step of $0.5$ fs.

We also carried out Born-Oppenheimer (BO) {\it ab initio} simulations in the NPT ensemble with the Qbox code \cite{qbox}, using a generalized gradient corrected approximation, PBE ~\cite{pbe}, for the exchange and correlation functional. We used norm conserving pseudopotentials and a plane-wave expansion of the Kohn-Sham orbitals with a kinetic-energy cut-off of $60$Ry. For all the systems investigated here we used a $\Gamma$ point sampling of the Brillouin zone. In our {\it ab initio} MD we substituted hydrogen with deuterium for computational convenience, so as to use a larger time step (5 a.u.) when integrating the Newton equations of motion. 

\section{Calculation of methane melting line}

Hysteresis effects near a solid liquid transition hamper an accurate estimate of melting temperatures from heat-until-melt MD runs. 
We carried out the calculation of the methane melting line by adopting two complementary techniques. 
One is based on the direct simulation of coexisting liquid and solid phases at given thermodynamic conditions, and we call it  two-phase coexistence method~\cite{twophase94}.  
The second approach consists of calculating the free energies of the solid and the liquid, and in obtaining the temperature of coexistence of the two phases by determining when their free energies are equal. Once a point on the coexistence line is known, it is then possible to find the whole coexistence 
curve by integrating the Clapeyron equation~\cite{gibbs}, i.e. by varying the pressure and 
temperature in such a way as to maintain equal the chemical potentials of the liquid and the solid: 
$\mu_l=\mu_s$.


The two-phase simulations were performed using periodic boundary conditions, with supercells containing $3456$ molecules, half of them prepared in the solid {\it fcc} phase and half of them prepared in the liquid phase, at pressures  of $0.01$, $0.5$ and $1.$ GPa. 
The liquid and solid phases were initially equilibrated separately in NPT ensemble runs
(constant number of particles, pressure and temperature) with fixed cell shape and 
then the two systems were merged together in a single supercell with a solid-liquid 
interface (we call the direction perpendicular  to the interface {\it z} axis ).

\begin{figure}
[ptb]
\vskip 0.5cm
\begin{center}
\includegraphics[height=2.in,width=3.in]{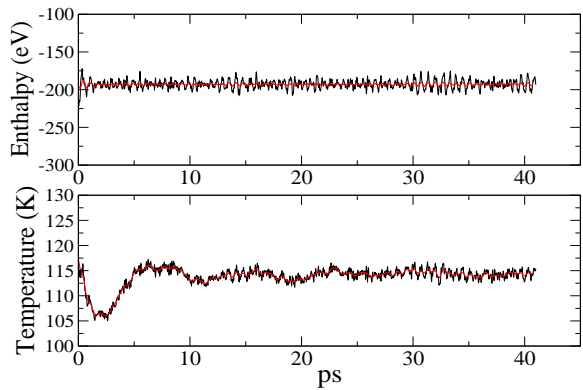}
\caption{Variation of enthalpy (upper panel) and temperature (lower panel) as a function of simulation time for a solid-liquid interface simulated in the (NPH) ensemble at $0.01$ GPa. The solid lines are the running averages over a $0.5$ps time interval.}
\label{temp_ent}
\end{center}
\end{figure}

We performed MD simulations of the coupled solid/liquid systems in the NPH ensemble (constant number of particles, 
pressure and enthalpy). During the simulation the pressure is applied isotropically while the cell angles cannot deform. If the initial pressure and temperature are sufficiently close to 
the conditions at which the solid and the liquid coexist, the temperature equilibrates to the value on the coexistence line, e.g. to the melting temperature, otherwise one phase prevails over the other. This approach differs from the one used in ref \cite{bonev,alfe}, 
where coexisting phases are simulated in the  NPT ensemble at different values of (P,T),  and for each run one determines the phase that is most stable under those conditions. 
When carrying out simulations in the NPH ensemble, if a reasonable choice of the initial $P-T$ conditions can be made, a single MD run suffices to determine the melting temperature. 
However, MD cells much longer than those adopted in NPT simulations are usually required.
As an example, we report a plot of the temperature and enthalpy at $0.01$ GPa in Fig.~\ref{temp_ent}.

In order to estimate finite size effects on our results, we carried out simulations with different cell sizes. The size of the simulated system may affect the values of the computed melting temperature in two ways: by the extension of the solid-liquid interface and, perhaps most importantly, by the size of the system in the direction orthogonal to the interface (the $z$ direction in our case). 
During the equilibration process in the NPH ensemble, solidification of the part of the system originally prepared as a liquid, and melting of the part originally prepared as a solid may occur, until a solid-liquid interface is stabilized. 
One of the two phases may take over, if the system size in the $z$ direction is not large enough, making it impossible to stabilize a two-phase system.
In table \ref{table1} we summarize the simulation results obtained with different supercell sizes. 
In our calculations, we first fixed the length of the cell in the direction orthogonal to the interface, and verified that changes in the extension of the solid-liquid interface does not substantially affect the computed value of the melting temperature: simulations performed with the three supercells $6\times6\times24$, $5\times5\times24$, $4\times4\times24$ give the same melting temperature with an error of less than $1$ K.
As expected, size effects along the $z$ direction are more dramatic. We considered the same geometry along the interface as indicated above (e.g. $6\times 6$ and $5\times5$ ) and we halved the length along the $z$ direction ($1726$ molecules  with $6\times6\times12$ supercell and $1200$ molecules with $5\times5\times12$ supercell ):  in this case  
the system completely transforms into one phase and no phase coexistence 
can be reached.
\begin{table}[t]%
\begin{tabular}
{cccc}\hline
molecules& supercell & enthalpy/molecule & $T_{melt}$ \\ 
& & $eV$ & $K$ \\
\hline
$3452$ & $6\times6\times24$ & $0.060$ & $114.3$ \\
$2400$ & $5\times5\times24$ & $0.056$ & $113.2$ \\
$1536$ & $4\times4\times24$ & $0.061$ & $113.3$ \\

$1726$ & $6\times6\times12$ & {\it no coexistence}  & \\
$1200$ & $5\times5\times12$ &  {\it no coexistence} & \\
\end{tabular}

\caption{Melting temperature of methane and enthalpy per molecule at $P=0.01GPa$ as obtained using different supercell sizes in (NPH) simulations}
\label{table1}%
\end{table}

As a validation of the results obtained with the two-phase method, we 
also computed the solid-liquid  boundary by a direct calculation of free 
energies. 
Using thermodynamic integration~\cite{frenkelti}, we can compute the Gibbs free energy difference (${\cal G}$) of a system with Hamiltonians $H^I$, and that of a reference system with Hamiltonian $H^{II}$, 
for which the melting curve is known by:
\begin{equation}
\label{free1}
{\cal G}^I - {\cal G}^{II} =  \int_{0}^{1} d\lambda \langle H^I- H^{II} \rangle_{\lambda}.
\end{equation}
The ensemble average in the integrand of Eq. (\ref{free1}) 
is evaluated by performing  MD simulations in the constant pressure and constant temperature
(NPT) ensemble, for a system with Hamiltonian $H(\lambda)=\lambda H^I - (1-\lambda) H^{II}$,
at different values of $\lambda$. 
As a reference for our thermodynamic integration we chose a Lennard-Jones system ($\varepsilon_{LJ} = 0.010323577$ eV and $\sigma_{LJ}=3.405$), which has the same equilibrium crystalline phase as methane, and for which the melting curve was accurately determined in previous calculations~\cite{kofke95}.
Free energy differences $\Delta {\cal G}$ 
were computed for the coexistence conditions of the 
reference system, both for the crystalline and the liquid phase. In particular we performed  
NPT runs of about 50 ps, for supercells containing 108 CH$_4$ molecules, and we considered 20 values of $\lambda$ for each phase. Size effects were tested by repeating simulations for a system containing 500 CH$_4$ molecules; the results for the melting temperature differed from those obtained with 108 molecules  by less than 1$\%$. 
Once two $\Delta {\cal G}$ values were obtained, we performed simulations for each phase, 
using the reversible scaling method~\cite{dekoning99,dekoning01} to determine the 
coexistence temperature, at a given pressure. 

This procedure has been applied to find the melting temperature at P=0, and P=0.5 Gpa,
yielding T$_c$=113 and 243 K, respectively, in excellent agreement with phase-coexistence
calculations (see Fig.~2).


We can now use one of the points on the melting line calculated with the previous methods, 
as the initial condition for integrating the Clapeyron equation.   
We start from the Gibbs-Duhem equation, 
\begin{equation}
d(\beta \mu)= hd\beta + \beta v dP,
\end{equation}
 where $\beta=1/kT$, $h$ and $v$ are the molar enthalpy and volume respectively, $P$ the pressure
We then write the Clapeyron condition $\mu_s=\mu_l$  as a first-order ordinary differential equation 
\begin{equation}
\label{clapeiron}
\frac{dP}{d\beta}= -\frac{\Delta h}{\beta \Delta v}
\end{equation}
where $\Delta h= h_l - h_s$ is the difference in molar enthalpies between 
the coexisting phases, while $\Delta v = v_l - v_s$ is the difference in volume; 
the derivative is evaluated along the coexistence line $\sigma$. 

We solved Eq. (\ref{clapeiron}) using a predictor-corrector formula;   
simultaneously performing two NPT simulations for the liquid and the solid and 
then evaluating  $\Delta h= h_l - h_s$ and $\Delta v = v_l - v_s$. 
We verified possible size effect using two different supercells of $108$ and $500$ molecules .
First we checked that the solution of Eq.~\ref{clapeiron} is independent  
on the initial conditions, i.e. the choice of different points on the melting line as initial states does not affect the results obtained for the melting line.
We also  verified possible size effect using two different supercells of $108$ and $500$. For temperature $T<350$ K size effects observed when solving Eq. (\ref{clapeiron}) are less than $1$\% and the supercells with $108$ and $500$ molecules give the same results. For $T>350$ the error on the melting temperature is about $5$\%, when using 108 molecules. 

\begin{figure}
[ptb]
\vskip 0.5cm
\begin{center}
\includegraphics[height=2.5841in,width=3.2396in]{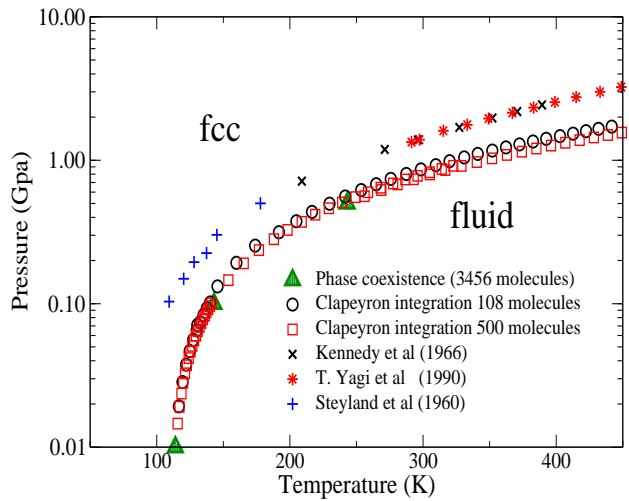}
\caption{Experimental and calculated melting line for the {\it fcc}-liquid phase transition. The experimental value are extracted from ~\cite{kennedy} (cross) ,~\cite{yagi} (stars),\cite{steyland}(plus). Calculated melting points are obtained with phase coexistence method (triangle) and integration of the Clapeyron equation \ref{clapeiron} (open circles and open squares respectively). We did not include the points calculated with free energy integration, because they  superimpose to the ones given by the phase coexistence method.    } 
\label{melting}
\end{center}
\end{figure}

All the results obtained for the methane melting line are summarized in Fig.~\ref{melting} together with the experimental values from ~\cite{kennedy,yagi, steyland} . The melting line obtained 
solving the Clapeyron equation is almost identical to that given by the 
phase-coexistence method and by free energy calculations.


The overall agreement between computed and measured melting temperatures is satisfactory, with a rigid shift of approximately $28$ K between the experimental and theoretical curves below  $200$K. Above $200$ K the slope of the theoretical and experimental melting curves differ by about a factor of $4$, making the agreement qualitative. The overall good agreement indicates that the forcefield used in our simulations is reasonably accurate in the description of methane-methane interactions at moderate pressures. We use the same force filed in the next section to investigate solid-solid phase transitions.
 
\section{Solid-solid phase transition in methane at moderate pressures}

\begin{figure}
[ptb]
\vskip 0.5cm
\begin{center}
\includegraphics[height=2.in,width=3.1in]{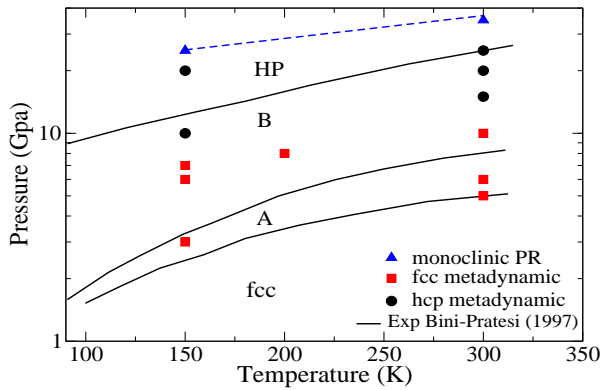}
\caption{Phase diagram of the solid methane. Experimental phase boundaries (solid line) are  from ~\cite{Bini}. Squares correspond to the  points   where the system does not evolve from an fcc during a metadynamics simulation. Circles represent  points where an fcc-hcp transition occurs during  metadynamics run.
A transition from hcp to a monoclinic crystal was observed with  Parrinello-Rahman simulations (triangles) on the hcp structure.The  dashed line is a tentative phase boundary between the hcp and the monoclinic phase}       
\label{final_solid}
\end{center}
\end{figure}

In order to predict the crystal structure of the higher pressure crystalline  phase of methane, 
we used  the metadynamics technique. Metadynamics~\cite{metadynamics} is a method for exploring free energy surfaces as a function
of selected collective coordinates, e.g. simulation cell edges ~\cite{metacell,natmat2006,prb2007}. This approach has been applied successfully
to several systems including, the crystal structure prediction of organic molecules~\cite{benzene,Karamertzanis08}.
Our simulations were performed with supercells of $108$ and $168$ molecules.

We first considered the compression of a {\it fcc} crystal ($108$ molecules) at $10$ GPa, and   
$150$ K and $300$ K. For both temperatures a transition to a {\it hcp} structure with four molecule per unit cell was observed, 
in which the molecules freely rotate in their final configuration. We did not find any evidence of 
intermediate stable configurations, nor we observed sizable jumps in the 
volume or in enthalpy of the system during the transition.
As discussed in ~\cite{prb2007,ddprl2008}, the analysis of the eigenvectors and eigenvalues of the Hessian matrix (see table (\ref{hesstab}) allows for a detailed description of the dominant crystal deformations occurring during the {\it fcc-hcp} transition.

\begin{figure}
[ptb]
\vskip 0.5cm
\includegraphics[
height=2.5in,
width=3.2in
]
{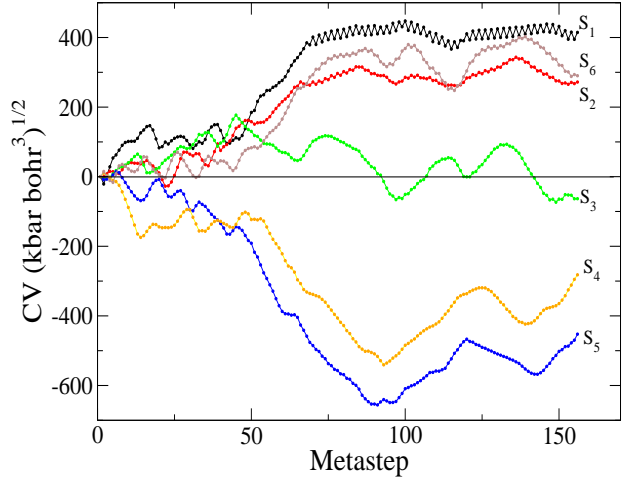}
\caption{Evolution of the metadynamics collective variable during the transition at  150 K and  10 GPa from fcc to hcp}
\label{evolution}
\vspace{0.4cm}

\end{figure}

The projection of the eigenvectors of the Hessian matrix on the six independent components of the cell edge matrix {\bf h}
(arranged in a six-component vector $(h_{11},h_{22},h_{33},h_{12},h_{13},h_{23})$), are related to the
elastic modes of the initial crystalline phase and the eigenvalues are related to the elastic moduli, i.e. the 
curvature of the free energy along the corresponding eigenvectors.
For example, the largest eigenvalue corresponds to the eigenvector $s_6$ in Tab.(~\ref{hesstab}) 
that points along the (111000) direction, and this is associated to a hydrostatic volume contraction/expansion.
In this case $s_1$ and $s_2$ are two equivalent orthorhombic deformations, which involve 
a compression/dilatation along pair of axes, while $s_3$, $s_4$ and $s_5$ are shear modes 
involving changes of the off-diagonal components of the Hessian matrix.
In  Fig.(\ref{evolution}) we report the evolution of the different modes  during a metadynamics simulation: the only inactive mode is $S_3$, confirming that a complex, concerted  rearrangement of the atoms take place during the {\it fcc-hcp} transition. 

Metadynamics simulations were performed also at $5$ GPa and $8$ GPa, but no additional stable crystal structures
 were found.
Based on a possible similarity between CH$_4$ and CF$_4$, Hirai {\sl et al.}~\cite{hirai} suggested 
that a phase A exists, with a cubic structure with $21$ molecule per cell, similar to a structure 
found for carbon fluoride.
In order to test this hypothesis, we performed a metadynamics simulation at 8 GPa 
and 300 K with 168 molecules/supercell, that is commensurate with  21 molecules per unit cell. 
In this case we observed a transition to a defective hcp structure and no stable structure 
corresponding to that of CF$_4$.


\begin{table}
\caption{\label{hesstab}Eigenvalues and the corresponding eigenvectors
for the Hessian matrix for methane metadynamic simulation at $150$K and $10$GPa }
\begin{tabular}{ccccccc}
\hline \hline
$Eigenvalue$ & $s1$ & $s2$ & $s1$ & $s3$ $s4$ & $s5$ & $s6$ \\
$(\times 10^3)$& $2.772$   &   $2.876$  &    $5.304$  &    $5.432$  &    $5.799$  &   $32.493$\\

\hline    
$Eigenvector$& & & & & &  \\
$h_{11}$&     $0.171$ & $0.793$ & $0.000$ & $0.017$ & $0.003$ & $0.583$\\
$h_{22}$& $0.595$ &  $-0.554$ & $0.048$ & $0.042$ & $0.014$ & $0.577$\\
$h_{33}$& $-0.779$ & $-0.250$ & $-0.051$ & $-0.050$ & $-0.007$ & $0.570$\\
$h_{12}$& $-0.053$ & $0.000$  & $-0.046$ & $0.664$ & $0.744$ & $-0.008$ \\

$h_{13}$& $0.040$ & $0.004$ & $0.003$ & $-0.743$ & $0.667$ & $0.000$ \\
$h_{23}$& $-0.071$ & $0.014$ & $0.996$ & $0.028$ & $0.031$ & $0.001$ \\

\hline \hline
\end{tabular}
\vspace{0.5cm}
\end{table}


\vspace{0.5cm}

\begin{figure}
[ptb]
\begin{center}
\includegraphics[
height=2.5841in,
width=3.2396in
]
{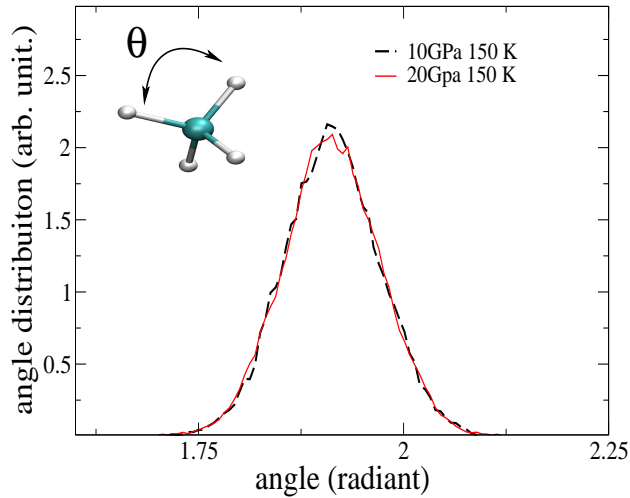}
\caption{Distribution of the intramolecular C-H, C-H angle calculated from {\it {\it ab initio}} molecular dynamics simulations at  10 GPa (dashed dotted  line),  20 GPa (solid line) and  150 K. }
\label{classicalangle}
\end{center}
\end{figure}

In order to explore the validity of the rigid molecule approximation used here, 
we performed BOMD simulations to test a posteriori our approximation. In particular, we carried out BOMD simulations starting from a {\it hcp} phase at  10 GPa and  20 GPa. After a short simulation of $\sim$1 ps, at both pressure we did not observe any significant distortions in the shape of the charge densities localized on 
the molecules,  and each CH$_4$ does behave as a rigid body. 
In Fig. (\ref{classicalangle}) we report the distribution of the intra-molecular 
angle between C-H bond. The angle distribution has the same profile both at 10 GPa and 20 GPa and is centered around the fixed value $\theta=1.9102$ rad used in our classical MD simulations.  
  
\begin{figure}
[ptb]
\vskip 0.5cm
\begin{center}
\includegraphics[
height=2.5in,
width=3.6in
]
{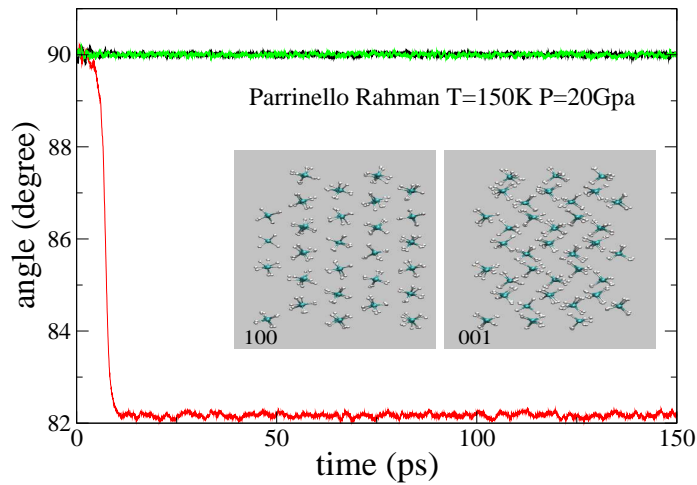}
\vspace{0.2cm}

\caption{Cell angle as a function of the simulation time. The Parinello-Rahman simulation was performed at 20 GPa and 150 K, starting from an hcp crystal structure with four molecule per unit cell, in an orthorhombic 
supercell containing 108 molecules.
 Inset: a snapshot of the final carbon position from the $[100]$ and $[001]$ directions.}
\label{angle}
\end{center}
\end{figure}

Starting from the {\it hcp} structure, we performed a series of constant pressure Parinello-Rahman  simulations for  different values of the pressure. 
At the temperature of $150$K the {\it hcp} structure with freely rotating molecules is stable up to $20GPa$, 
when deformation of one of the cell angle occurs (Fig.(\ref{angle}))
The hcp cell with four molecules per unit cell transforms into a monoclinic 
structure (P21/c) with four molecule per unit cell. At $20$GPa the molecules stop rotating and a denser crystal structure is stabilized. 
The ``freezing'' of the hydrogen atoms is accompanied by a deformation of one of the cell angles as clearly shown in Fig.(\ref{angle}), where the angles are plotted as a function of the simulation time.  
A snapshot of the carbon positions in the final configuration at $20$GPa, is shown in the inset of Fig.~\ref{angle}  along the $[100]$ and $[001]$ directions.


\section{Conclusion}

We have presented results on the methane melting line, in 
satisfactory agreement with existing measurements and discussed in 
detail the comparison between several different computational 
techniques. These results also validate the empirical potential~\cite{potenziale}
in the thermodynamic regime considered in this work.

In addition we propose two crystal structures for methane at moderate 
pressure and we provide insight into the transition mechanisms between such crystalline phases.
The results from metadynamics calculations and Parrinello-Rahman  simulations are  summarized in Fig. \ref{final_solid}, together with the experimental boundary lines obtained from ~\cite{Bini}. 
In our simulations we find that under pressure, at temperatures between 150 and 300 K,
the fcc crystals transforms  into a hcp structure with 4 molecules 
per unit cell (B-phase) and then into a monoclinic (4 molecules per cell) high pressure phase. 
These results are consistent with those of IR and Raman investigations by Bini 
{\it et al.}~\cite{Bini}, although we did not find any A phase, intermediate 
between the fcc and B phases. In addition, the estimated solid/solid 
phase boundaries are at higher pressure in our calculations than those 
experimentally inferred from spectroscopic experiments. The absence of an 
A phase is not entirely surprising as the A-B transition has been found 
to be very sluggish in all experiments reported to date, and it 
may be accompanied by a complex atomic rearrangements not commensurate 
with the MD cells used in our simulations. We did not find any evidence 
of a phase with 21 molecules per unit cell proposed in Ref.~\cite{hirai} by 
analogy with the CF$_4$ crystal and our results do not seem to agree with the 
interpretation of the experimental findings of Refs.~\cite{nakahata,umemoto}.
While additional work is required to understand in detail the solid 
phases of methane, our simulations represent a useful set of qualitative 
results, consistent with available experimental spectroscopic evidence, 
on which more refined {\it ab initio} MD simulations can be based upon.

Further work is in progress to study the methane phase diagram in a 
regime where molecules dissociate, using first principle techniques.

We ackwnowledge helpful correspondance with  Roberto Bini and O. Andreussi. 
We thank R.Hemley and  M. Somayazulu for useful discussions.
This work was supported by Shell Corporation.



\bibliography{bibliografia}

\end{document}